\documentclass[twocolumn,showpacs,preprintnumbers,amsmath,amssymb]{revtex4}

\usepackage{graphicx}
\usepackage{bm}

\begin{document}

\preprint{APS/123-QED}

\title{NON-CLASSICALITY INDICATORS FOR ENTANGLED AND SQUEEZED NUMBER STATES}

\author{Siamak Khademi}
\email{siamakkhademi@yahoo.com} \email{skhademi@znu.ac.ir}
\author{Elham Motaghi}
\affiliation{Department of Physics, University of Zanjan, ZNU,
Zanjan, Iran.}%
\author{Parvin Sadeghi}
\affiliation{ Marand Faculty of Engineering, University of Tabriz, Tabriz, Iran.}%

\date{\today}

\begin{abstract}
Many non-classicality indicators are used to measure quantum effects of different systems. Kenfack's and
Sadeghi's non-classicality indicators are introduced in terms of the amount of Wigner function's negativities
and interferences in phase space quantum mechanics, respectively. They are, effectively, applied for some real
distribution functions. In this paper, we compared these non-classicality indicators for the entangled state of
photonic number states in the Wigner, Husimi and Rivier representations. It is shown that for a two-level
entangled state, Sadeghi's indicator has more benefits with respect to the  Kenfack's indicator. For the
two-level entangled state, we show a correspondence between the Sadeghi's non-classicality indicator and the
Von Neumann entropy. It is also shown that for the superposition of squeezed number state the Sadeghi's (and no
Kenfack's) non-classicality indicators is sensible with respect to the squeezing parameter for a superposition
of squeezed number states
\end{abstract}

\maketitle

\section{INTRODUCTION}
The phase space formulation of quantum mechanics, presented by
Wigner in 1932 \cite{Wigner1932}, offers a classical like
formulation. He proposed Wigner distribution functions (WDFs),
which in general, were real and non-positive
\cite{Lee1995,Hillery1984}. Some authors are believed the
negativity of Wigner distribution functions has indicated to some
nonclassicality phenomena \cite{Nogues2000,Rigas2011}. Kenfack and
\.{Z}yczkowski invent a nonclassicality indicator which measured
the negativity of Wigner distribution function
\cite{Kenfack2004}. Their nonclassicality indicator has been applied
for different quantum systems
\cite{Zhen2013,Talebi2012,Chatterjee2012}. As well as WDF, there
is many distribution functions in different phase space representations e:.g.: Kirkwood
distribution function (KDF) \cite{Kirkwood1933}, Husimi
distribution function (HDF) \cite{Husimi1940}, Rivier
distribution function (RDF) \cite{Rivier1951} and so on
\cite{Glauber1963,GlauberBook,Chaturvedi1977,Drummond2003}. Among
them the HDF is positive definite and has no negativity at all
\cite{Lee1995}. The expectation values of physical quantities are
assumed to be independent of different phase space representations,
therefore all distribution functions should be, in general,
equivalent \cite{Lee1995,Hillery1984}. Application of a specific phase space representation
for special systems, have their benefit in
calculation \cite{Lee1995,Hillery1984}. The Kenfack and
\.{Z}yczkowski's nonclassicality indicator is just defined in
terms of WDF \cite{Kenfack2004}. In general it doesn't suitable
for other real distribution functions \cite{Sadeghi2010}. It
specially vanishes for the positive definite distribution
functions like HDFs. The equivalence of distribution functions
lead authors to ask, "if the negativity of WDFs indicate to the
physical quantum effects, what about them in other distribution
functions?".

Clearly the Kenfack and \.{Z}yczkowski's
nonclassicality indicator (which doesn't vanish for WDFs)
vanishes for HDF. To remove this inconsistency another
nonclassicality indicator is introduced by Sadeghi {\it et al.}
\cite{Sadeghi2010} based on the interference of quantum states.
It is shown that the Sadeghi's nonclassicality indicator works
properly for some real distribution functions, e.g. Husimi, and
Rivier as well as Wigner \cite{Sadeghi2010}. In this paper these
nonclassicality indicators are applied for the entangled and a superposition state
of squeezed number states to measure the
amount of entanglement and squeezing parameter. As an example;
the entanglement of a two-level and squeezed number states are
investigated to compare the benefits of two indicators. In the next
section we have a brief review about the Kenfack and \.{Z}yczkowski,
and Sadeghi's nonclassicality indicators. Section 3 and 4 belong
to the calculation and comparison of nonclassicality indicators
for the entangled photonic and squeezed number states,
respectively. Section 5 is devoted to the conclusions.

\section{Nonclassicality indicators}
The Wigner distribution function for the state $|\psi \rangle$ is
defined by \cite{Wigner1932}
\begin{equation}
W(q,p)=\int_{-\infty}^\infty dx \langle q+\frac{x}{2}|\psi
\rangle \langle \psi|q-\frac{x}{2}\rangle,\label{eq1}
\end{equation}
 Kenfack and \.{Z}yczkowski define a
nonclassicality indicator which depends on the amount of
negativity in Wigner distribution function as
follows\cite{Kenfack2004}:

\begin{eqnarray}
\delta_w &=&\int_{-\infty}^\infty [|W(q,p)|-W(q,p)]dq dp\nonumber\\
&=&\int_{-\infty}^\infty |W(q,p)|dq dp-1,\label{eq2}
\end{eqnarray}

The nonclassicality indicator $\delta _w$ is just defined in terms
of Wigner distribution function and is equal to zero for coherent
and squeezed vacuum states, for which the corresponding Wigner
distribution functions are non-negative. It is also not sensitive to
the squeezing parameter, because the negativity of Wigner
distribution function is conserved due to squeezing operation.
This nonclassicality indicator doesn't work for other
distribution functions, like Husimi or Rivier, properly
\cite{Sadeghi2010}. Another nonclassicality indicator is
introduced by Sadeghi {\it et al.} according to the interference
of quantum states in phase space \cite{Sadeghi2010}. This
nonclassicality indicator is defined by
 \begin{equation}
\eta  = \frac{{\sum\limits_{ij} {\int_{ - \infty }^\infty
{[|f_{ij} | - f_{ij} ]} } dqdp}}{{\sum\limits_{ij} {\int_{ -
\infty }^\infty {[|f_{ij} | + f_{ij} ]}}dqdp}}.\label{eq3}
\end{equation}
where the phase space distribution function for a superposition
of quantum states $\psi=\psi_1+\psi_2$, is divided into four parts
$f=f_{11} (\psi_1 )+f_{22} (\psi_2 )+f_{12} (\psi_1,\psi_2^*
)+f_{21} (\psi_2,\psi_1^*)$ (the last two parts are interference
terms). In Eq.~(\ref{eq3}), $f_{ij}$'s are different parts of
distribution function and sum is over all parts
(interference and non-interference terms). This nonclassicality
indicator is bounded between $0$ and $1$. This indicator is also
applied for some different real distribution functions, e.g.
Wigner, Husimi and Rivier. It is shown that the behavior of the
nonclassicality indicator $\eta$ doesn't depend on the phase space
representations for the Schr\"{o}dinger cat and thermal states
\cite{Sadeghi2010}.

\section{Nonclassicality indicators for entangled states}
\begin{figure}[htb]
\begin{center}
\includegraphics[height=5cm]{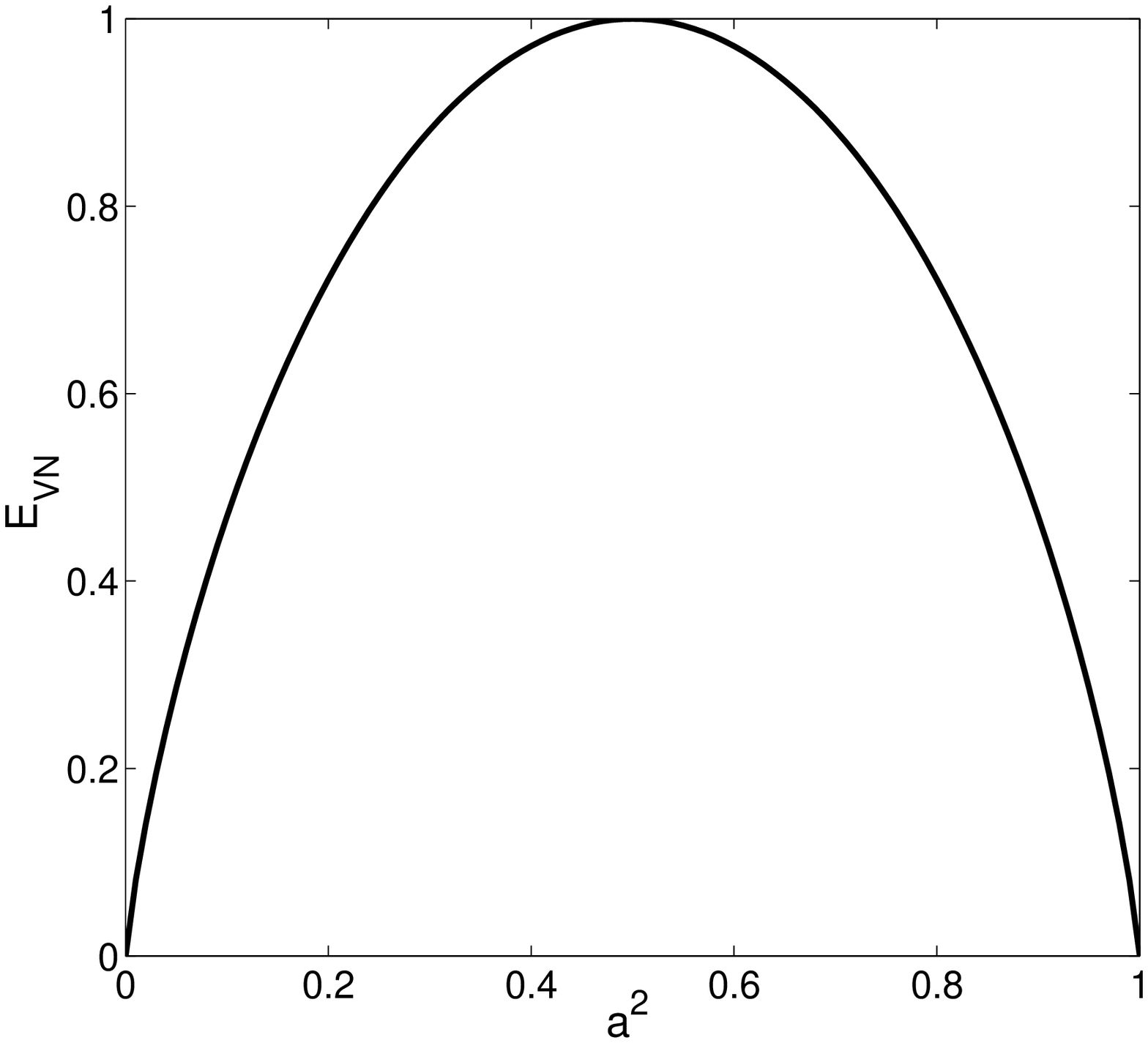}
\includegraphics[height=5cm]{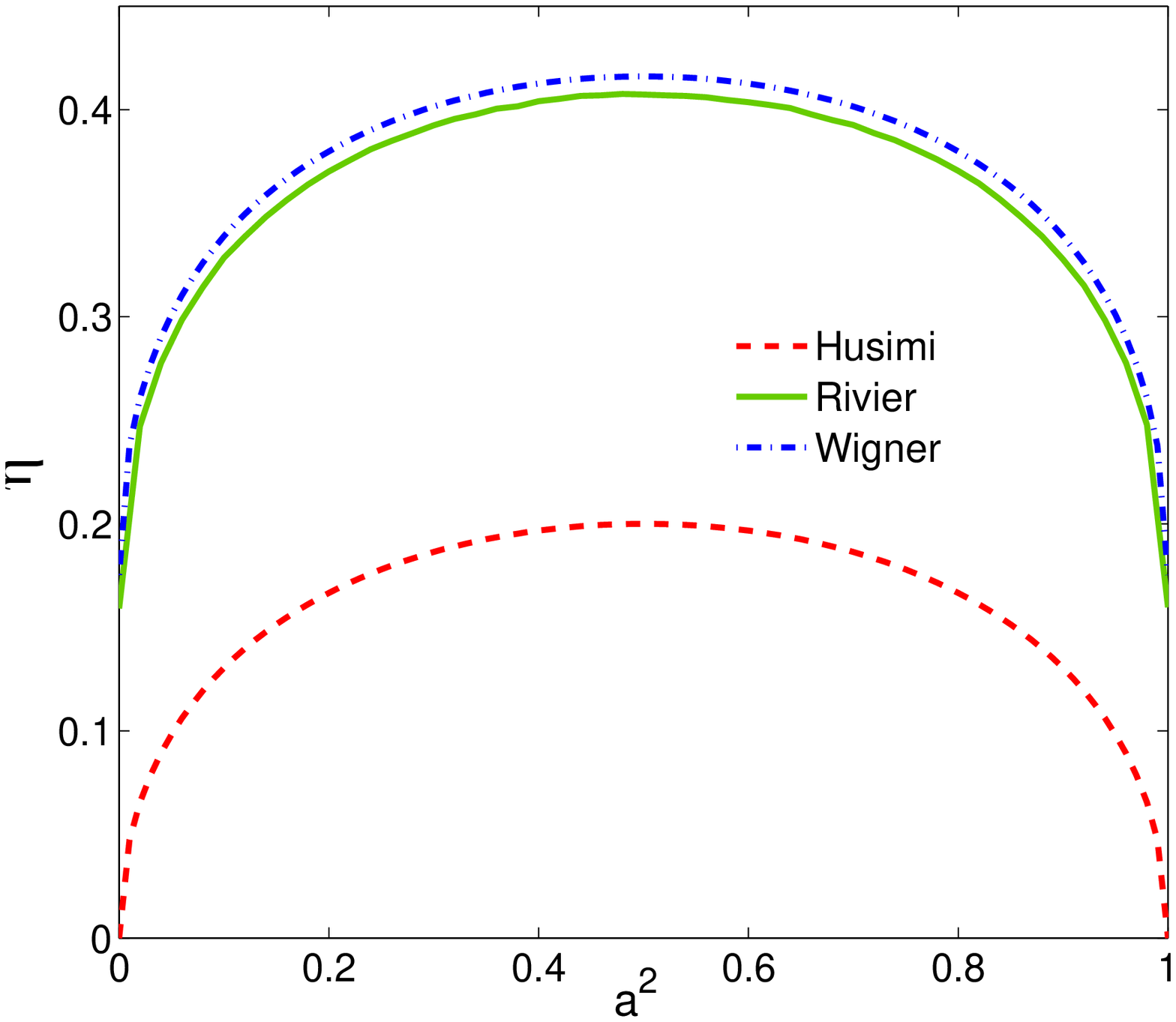}
\caption{(Color online) Left)The Von Neumann entropy, $E_{VN}$, for an
entanglement state of first and second excited state of harmonic
oscillator vs. $a^2$. The parameter $a$ is a superposition state constructor and is supposed to be a real
number for simplicity. Right)The non-classicality indicator $\eta$ for
an entanglement state of first and second excited state of
harmonic oscillator vs. $a^2$ for Wigner, Husimi and Rivier
distribution function.}
\label{fig1}
\end{center}
\end{figure}

The entangled state between the ground and first excited number
states (or the first excited and second excited number states,
and so on) is given by $|\psi \rangle=a|0,1\rangle+(1-a^2
)^{1/2}|1,0\rangle$(or$|\psi \rangle=a|1,2\rangle+(1-a^2
)^{1/2}|2,1\rangle$,etc.). The parameter "$a$" is supposed to be a real
number for simplicity. The corresponding Wigner function is given
by
\begin{equation}
W(q,p)=W_{11}+W_{22}+W_{12}+W_{21},\label{eq4}
\end{equation}
where
\begin{equation}
W_{11}(q,p)=\frac{2a^2}{\pi ^2}(q_1 ^2+p_1 ^2-\frac{1}{2})
\exp[-q_1 ^2-p_1 ^2-q_2 ^2-p_2 ^2],\label{eq5}
\end{equation}
\begin{eqnarray}
W_{22}(q,p)&=&\frac{2(1-a^2)}{\pi ^2}(q_2 ^2+p_2 ^2-\frac{1}{2})
\nonumber\\
&& \exp[-q_1 ^2-p_1 ^2-q_2 ^2-p_2 ^2],\label{eq6}
\end{eqnarray}
\begin{eqnarray}
W_{12}(q,p)+W_{21}(q,p)&=&\frac{4a\sqrt(1-a^2)}{\pi ^2}(q_1
q_2+p_1
p_2) \nonumber\\
&&\exp[-q_1 ^2-p_1 ^2-q_2 ^2-p_2 ^2],\label{eq7}
\end{eqnarray}
where $q_1,q_2,p_1,p_2$ are the generalized coordinates and
momenta in phase space and the interference terms $W_{12}(q,p)=W_{21}(q,p)$. Clearly
the nonclassicality indicator has a constant values
$\delta_w=0.426$ for entangled between ground and first excited
states (and $\delta_w=0.653$ for the first and second excites
states), independent on the value of "$a$". It is also a constant
value for Rivier distribution function and vanishes for the
Husimi distribution function. Therefore this nonclassicality
indicator is not suitable to indicate the entanglement property
of this system and has not a correspondence with the Von Neumann
entropy for this system which is shown in Fig.~\ref{fig1}.

\begin{figure}[htb]
\begin{center}
\includegraphics[height=5cm]{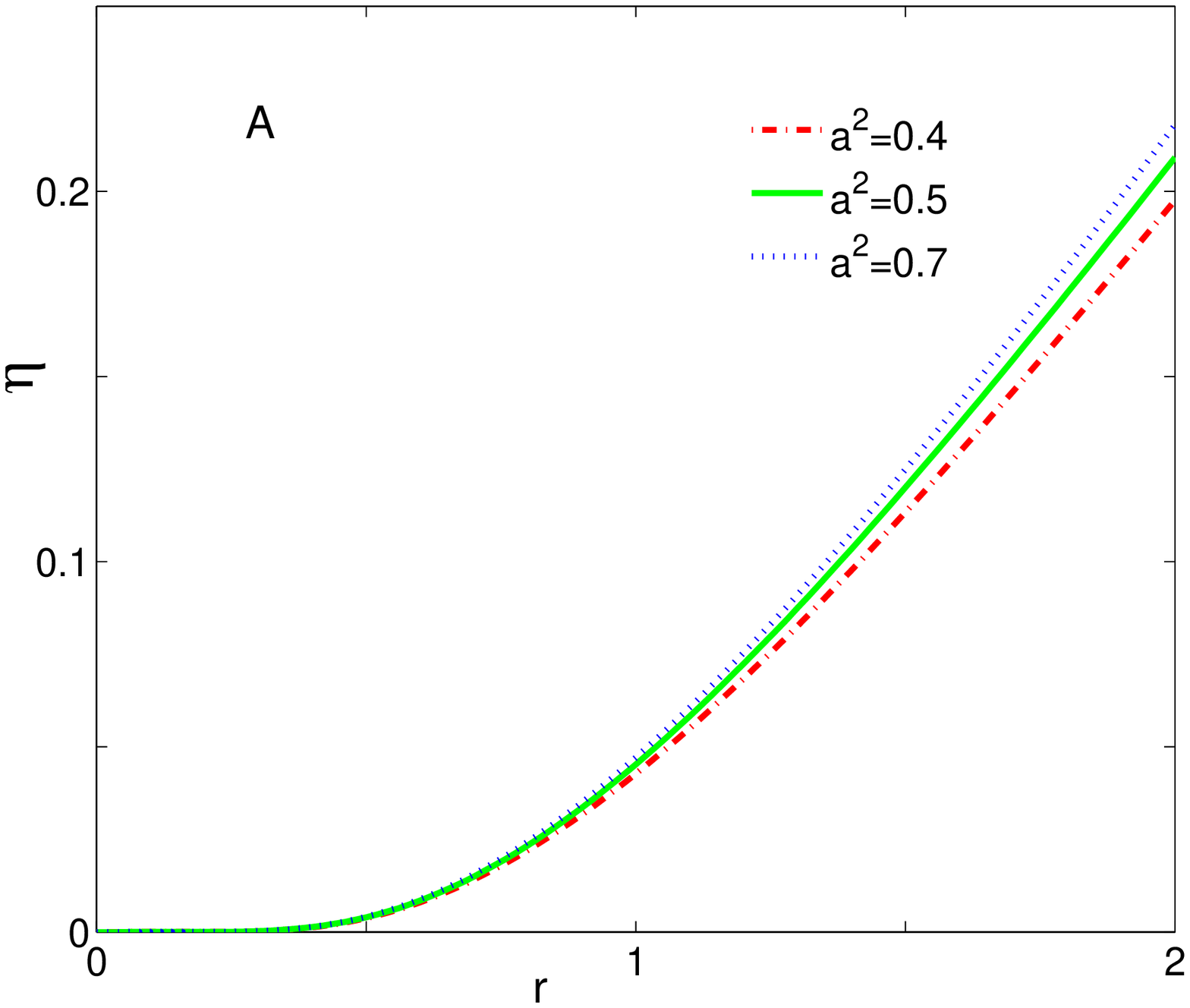}
\includegraphics[height=5cm]{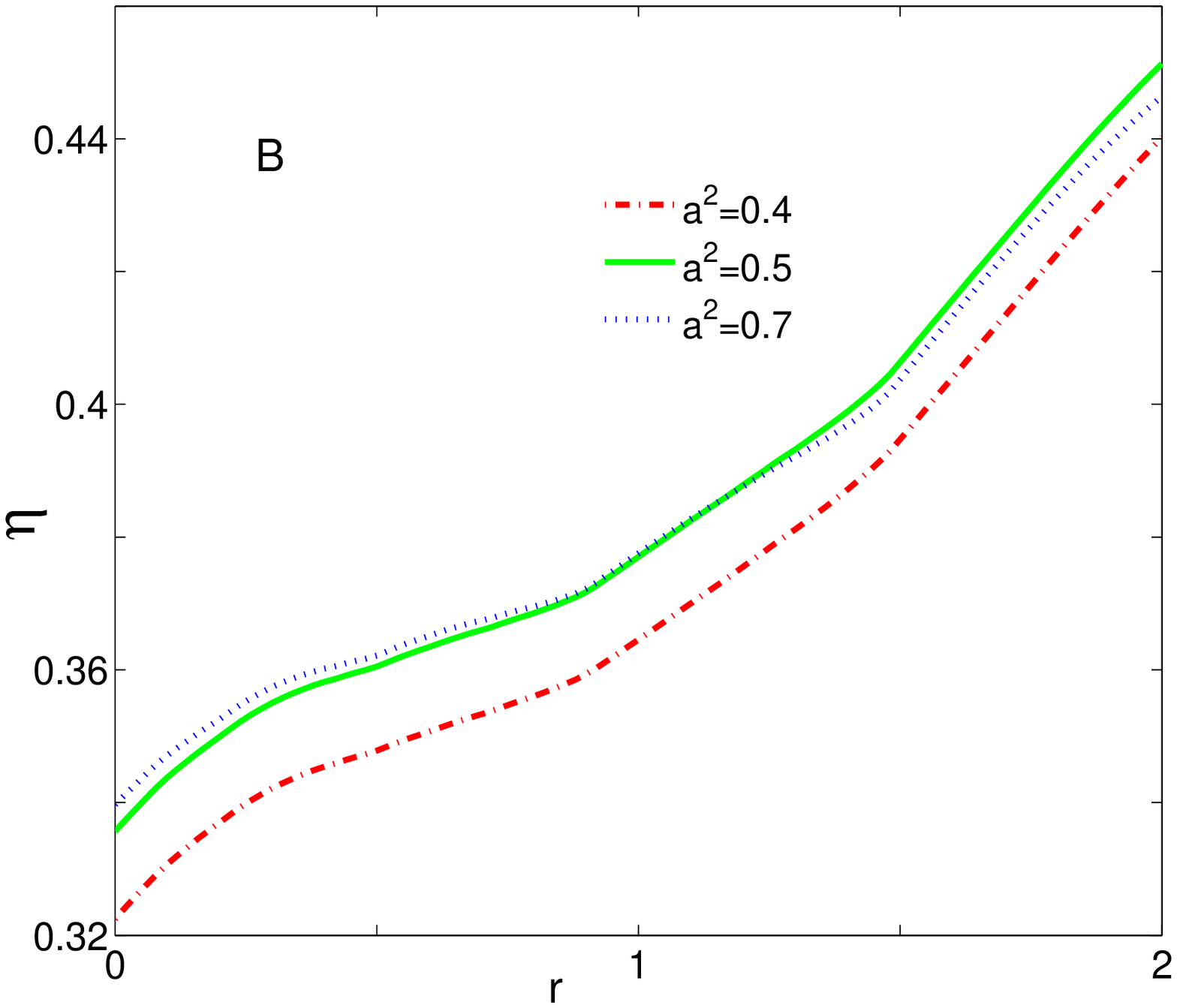}
\caption{(Color online) The non-classicality indicator $\eta$ for
squeezed states, A: the superposition of ground and squeezed
ground state of number state $| \psi_{00r}\rangle$; B: the
superposition of ground and squeezed first excited state $\psi_{01r}\rangle$ vs. $r$ for different values of
probability
amplitude in the Wigner representation.} \label{fig2}
\end{center}
\end{figure}

Now we investigate the behavior of nonclassicality indicator $\eta$ for
the same state in the Husimi, Rivier, and Wigner representation.
The results are plotted in Fig.~\ref{fig2}. The nonclassicality
indicator $\eta$ has a constant value in the end points $a=0,1$
\footnote{The negative values in the end points are due to the negativity of individual state $|0,1\rangle$
or $|1,0\rangle$, which is indicated to their quantum uncertainty
relation}. The nonclassicality indicator $\eta$ has also a maximum
at $a=\pm 1/2$  which is corresponding to the maximum
entanglement of the Bell states. This maximum point is clearly
independent of phase space representations, e.g. Wigner, Rivier
and Husimi, which are investigated in this paper. To show this
correspondence more clearly, it is suitable to compare the
nonclassicality indicator $\eta$ in different representations
which is plotted in Fig.~\ref{fig2} and the Von Neumann entropy
in Fig.~\ref{fig1}. Although the nonclassicality indicator has
different values for different distribution functions, its
behavior has a suitable correspondence with the Von Neumann
entropy. Application of this method for the entangled state of
first and second photonic excited state, and other entangled
states, is straight forward.
There is a correspondence between the Kenfack and \.{Z}yczkowski
nonclassicality indicator $\delta$, the Sadeghi's nonclassicality
indicator $\eta$ and the Von Neumann entropy for these entangled
states.

\section{Non-classicality indicators for the squeezed states}
The squeezed number states are another example from nonclassical
states. The Kenfack and \.{Z}yczkowski nonclassicality indicator
$\delta$ is not sensitive to the squeezing parameter, due
to the amount of negativity for the squeezed state is conserved in squeezing
operation. In this section we consider two superposition states:
\\1) The superposition of ground and squeezed ground number states $|
\psi_{00r}\rangle=(1-a^2 )|0\rangle+a|0,r\rangle$,
\\2) The superposition of ground
and squeezed first excited states $|\psi _{01r}\rangle=(1-a^2
)|0\rangle+a|1,r\rangle$,
where "$a$" and "$r$" are real probability
amplitude and squeezing parameter, respectively. Four parts
of Wigner function for $|\psi _{00r}\rangle$ are given by
\begin{equation}
W_{11}(q,p)=\frac{1-a^2}{\pi }\exp[-q^2-p^2],\label{eq8}
\end{equation}
\begin{equation}
W_{22}(q,p)=\frac{a^2}{\pi}\exp[-e^{2r}q^2-e^{-2r}p^2],\label{eq9}
\end{equation}
and
\begin{eqnarray}
W_{12}(q,p)&+&W_{21}(q,p)=\frac{2a\sqrt(2*(1-a^2))}{\pi\sqrt(1+e^{2r})}\nonumber\\
&&\cos(\frac{2qp(e^{2r}-1)}{1+e^{2r}})\nonumber\\
&\times &
\exp[\frac{r}{2}-\frac{2e^{2r}q^2}{1+e^{2r}}-\frac{p^2}{1+e^{2r}}].\label{eq10}
\end{eqnarray}
Clearly, the nonclassicality indicator $\eta$ for $|\psi
_{00r}\rangle$ state is plotted in Fig. 3A. The nonclassicality
indicator $\eta$ is an increasing function vs. the squeezing
parameter $r$. The similar result is obtained for $|\psi
_{01r}\rangle $ which is plotted in Fig. 3B. Therefore the squeezing
parameter is also a measurable quantity with Sadeghi's
non-classicality indicator.

\section{CONCLUSION}
The nonclassicality indicators  $\delta$ is applied just for
Wigner distribution function in phase space but the
non-classicality indicator$\eta$ is applicable for more real
distribution functions like; Husimi and Rivier. This
nonclassicality indicator is applied to indicate to the
entanglement which is one of the most important nonclassical
phenomena. For an example, the entangled states of two
eigenstates are used. For an entangled state consist of the
ground and the first excited state the nonclassicality indicators
$\delta$  has a constant value and therefore is not a suitable
indicator for entanglement. In the other hand, the
nonclassicality indicator  $\eta$ has a maximum for a Bell state
and has a correspondence with the Von Neumann entropy, for the
investigated system. Developing to the other entangled states,
e.g. the first and second excited states, is straight forward.
For these systems the nonclassicality indicator $\delta$  has also a
correspondence to nonclassicality indicator  $\eta$ and Von
Neumann entropy, and has a maximum according to the Bell state
which is independent of phase space representations. Also, the
nonclassicality indicator $\eta$  has more correspondence with the
squeezing parameter $r$.


\end{document}